# CasCIFF: A Cross-Domain Information Fusion Framework Tailored for Cascade Prediction in Social Networks

Hongjun Zhu[a,b]*, Shun Yuan[a,b], Xin Liu[a,b,c], Kuo Chen[a,b], Chaolong Jia[a,b] and Ying Qian[a,b]*

[a] School of Software Engineering, Chongqing University of Posts and Telecommunications, Chongqing 400065, China

[b] Chongqing Engineering Research Center of Software Quality Assurance, Testing and Assessment, Chongqing 400065, China

[c] School of Engineering, Computer and Mathematical Sciences, Auckland University of Technology, Auckland, 1010, New Zealand

E-mail: zhuhj@cqupt.edu.cn, s211231071@stu.cqupt.edu.cn, {liuxin, chenkuo, jiacl, qianying}@cqupt. edu.cn

*Corresponding author: Hongjun Zhu, Ying Qian.

**Abstract**

The explosive growth of the Internet has elevated social networks to a pivotal role in information propagation, reshaping conventional paradigms of information distribution. Utilizing the vast amount of data available online to model and predict this diffusion is of great importance in various fields. Existing approaches for information cascade prediction fall into three main categories: feature-driven methods, point process-based methods, and deep learning-based methods. Among them, deep learning-based methods, characterized by their superior learning and representation capabilities, mitigate the shortcomings inherent in the other methods.

However, current deep learning methods still face several persistent challenges. In particular, accurate representation of user attributes remains problematic due to factors such as fake followers and complex network configurations. Previous algorithms that focused on the sequential order of user activations often neglected the rich insights offered by activation timing. Furthermore, these techniques often fail to holistically integrate temporal and structural aspects, thus missing the nuanced propagation trends inherent in information cascades.

To address these issues, we propose the Cross-Domain Information Fusion Framework (CasCIFF), which is tailored for information cascade prediction. This framework exploits multi-hop neighborhood information to make user embeddings robust. When embedding cascades, the framework intentionally incorporates timestamps, endowing it with the ability to capture evolving patterns of information diffusion. In particular, CasCIFF seamlessly integrates the tasks of user classification and cascade prediction into a consolidated framework, thereby allowing the extraction of common features that prove useful for all tasks, a strategy anchored in the principles of multi-task learning. After extensive experiments conducted on publicly available datasets, the results demonstrate CasCIFF's superiority over established baseline methods in terms of prediction accuracy.

# 1. INTRODUCTION

In today's digital era, social networking platforms such as Weibo and Twitter have emerged as critical vehicles for information propagation. With the growing influence of such networks, the ability to predict the extent of information diffusion becomes desirable for applications including, but not limited to, media governance [1], advertising marketing [2], and scientific impact prediction [3]. As a result, this topic has attracted considerable attention from both the academic and industrial communities.

In the field of social networks, the spread of information is often conceptualized as an information cascade phenomenon. In this context, an information cascade describes both the pathways through which information spreads and the participants involved in such processes [4]. Contemporary research on information cascade prediction generally falls into two categories based on task specificity: (1) micro-level prediction, which focuses on predicting the next likely participant in a given information cascade [5-7]; and (2) macro-level prediction, which evaluates the size of information diffusion [4, 8, 9]. The present study is mainly anchored in the macro-level perspective, aiming at predicting the future growth of the cascade size.

Current methods for cascade prediction fall into three main categories [4]: feature-driven methods, point process-based methods, and deep learning-based methods. Feature-driven methods exploit cascade attributes such as content, timing, and structure to detect diffusion patterns [8]. Despite their advantages, these methods rely heavily on domain expertise and may underperform in unfamiliar scenarios [10]. Point process-based methods utilize models like the Poisson process [11] and the Hawkes process [12] to encapsulate the dynamics of information propagation. Despite their usefulness, these methods often oversimplify the intricate dynamics of information spread, potentially compromising prediction accuracy [4]. Deep learning-based methods have gained significant traction in the field of information propagation, with methods such as DeepHawkes [13], CasCN [14], CasFlow [15] marking notable advances in the prediction of information cascades.

Despite these methodological advances, the dominant models still have their own limitations.

(1) Representation of global user influence: Current evaluations of user influence are predominantly based on the number of followers, neglecting the qualitative aspects, leading to potential misevaluation due to the undue influence of fake followers [16], which, in turn, can distort the accuracy of prediction models.

(2) Representation of dynamic cascades: Information cascades are dynamic in nature and their sizes vary with not only diffusion patterns but also propagation speeds. However, previous cascade prediction algorithms largely concentrate on the chronological order of user activations, thereby neglecting the insights offered by action timing [17].

(3) Integration of multi-source data: Contemporary cascade prediction architectures tend to be reliant only on singular or specific feature sets, ignoring the nuances and correlations across diverse features [13], which results in the potential omission of valuable cross-domain insights [12].

In response to these limitations, we propose CasCIFF, an integrative framework tailored for information cascade prediction. This model seamlessly fuses the structural properties inherent in social networks with the temporal dynamics of information cascades.

Our main contributions include:

(1) We propose a novel strategy to represent user influence that is rooted in the global social

network structure. This approach mitigates the negative impact of fake followers by exploiting information from multi-hop neighboring nodes.

(2) We introduce a cascade representation that improves the efficiency of graph embedding. A normalized interaction timing is introduced into the adjacency matrix and linked to feature vectors, ensuring a comprehensive capture of propagation dynamics.

(3) We construct a spatiotemporal data fusion framework supported by multi-task learning, which integrates global user characteristics, local cascade attributes, and temporal dynamics to simultaneously address the dual challenges of cascade prediction and user classification.

(4) We conduct rigorous evaluations using real-world datasets. Our exhaustive experiments on three datasets demonstrate CasCIFF's superior performance over existing methods in terms of prediction accuracy. To facilitate further work in this area, the source code is available at https://github.com/XiaoYuan011/CasCIFF.

# 2 RELATED WORK

As mentioned above, existing methods for cascade prediction can be divided into three main categories: feature-driven methods, point process-based methods, and deep learning-based methods. In this section, we briefly review some of the research literature related to our work.

## 2.1 Feature-driven methods

Feature-driven methods aim to predict the magnitude of an information cascade by extracting and exploiting features inherent in cascades. Such features provide a cue for cascade prediction, including the number of nodes involved, the structure of the cascade tree, the timing of events, or the content associated with the cascade.

Several studies have highlighted the role of these features. For instance, Cheng et al. [8] emphasized the significance of temporal features in cascade prediction. Bakshy et al. [18] probed the correlation between follower count and tweet popularity, demonstrating the role of influential leaders in information propagation. Ashton Anderson [19] found a strong correlation between cascade popularity and network topology. Wu et al. [20] discussed the influence of content type on cascade popularity. Once these features are extracted, they are typically fed into machine learning or statistical models to complete the prediction task.

The primary advantage of feature-driven methods is their ability to leverage powerful machine learning to incorporate diverse information. However, the effectiveness of these techniques is highly dependent on the quality and relevance of the manually extracted features [21]. Additionally, these methods often rely on static feature representation and hence hardly capture the complex, dynamic nature of cascades. Moreover, the design of these features is often tailored to specific networks or datasets, which could compromise performance and transferability when applied to different scenarios [13].

## 2.2 Point process-based methods

A point process is characterized by a stochastic collection of points within a mathematical continuum, where each point represents an event occurring at a particular time[22]. From this perspective, Shen et al. [11] proposed a framework for cascade prediction, which is based on the assumption that information cascades follow a Poisson process, i.e. events occur continuously and independently at a constant mean rate. However, this assumption does not follow from the real-world information cascade, where the frequency of event occurrence depends on the historical processes [23].

In response to this limitation, the Hawkes process, a type of self-exciting process, has gained scientific recognition. The Hawkes process extends the Poisson process by allowing the event rate to be influenced by its historical trajectory, thereby reflecting complex dynamics. For example, Zhao et al. [24] utilized multivariate Hawkes processes to model information diffusion. Pinto et al. [25] further refined this model by integrating user interactions and temporal effects. Despite their success in certain domains, an inherent assumption of the Hawkes process - that each event increases the likelihood of upcoming events - may not always be true [26]. Such potential oversimplifications can lead to performance limitations [27-29].

## 2.3 Deep learning-based methods

Recently, methods based on deep learning have attracted considerable attention due to their impressive performance. Deep learning provides an end-to-end solution, where the quality of input data directly affects the performance of cascade prediction. To input cascade information, DeepCas [29] transformed the cascade graph into a user sequence using the random walk strategy, a technique proposed in DeepWalk [30]. Similarly, CasHAN [16], DeepCon and DeepStr [28] employed random walks to extract information, with the latter using second-order random walks to capture more information.

Despite the efficacy of random-walk sequences in elucidating network architecture, they often ignore the temporal order of user activations, a crucial factor in understanding cascade dynamics [13]. To reconcile this problem, many researchers use time-ordered sequences instead of random-walk sequences [5, 13, 31, 32]. While these sequences capture the temporal dynamics of the cascade, they may overlook contextual information, such as the network structure.

To address this issue, the time-ordered snapshot (TOS) method has gained traction for information capture [12, 15, 33]. The TOS method preserves both temporal order and network topology. However, its potential drawback is an omission of precise event timings or inter-event intervals, possibly leading to an oversight of dynamic trends. To overcome this limitation, techniques presented by CasTCN [34] and MUCas [10] advocate equal interval snapshots (EIS). Nevertheless, EIS are susceptible to loss of valuable detail if the chosen interval is too large, especially when significant events surge within a short period of time.

To mitigate this problem, AECasN [17] inputs the cascade graph and the corresponding retweet time as a whole. This method then generates an embedding to encapsulate the holistic cascade landscape. Unfortunately, it neglects the individuality of the user, which can influence the way of social interactions and message propagation [35].

In essence, user influence and susceptibility inherently determine user behavior and consequently cascade size. For this reason, methods such as HeDAN [36] and PFDID [31] integrate the information of both cascade graph and user individuality. However, these methods predominantly characterize user attributes based on 1-hop neighbor information, which potentially compromises the robustness of the representation, especially in the presence of fake followers within the network.

Overall, existing methods have been successful in cascade prediction for specific scenarios, while they also have limitations in capturing the complex, dynamic nature of information cascades, integrating both temporal and structural information, and considering user-specific characteristics.

# 3 PRELIMINARIES

This section introduces the necessary background and fundamental concepts that are critical for understanding the remainder of this paper. In this context, various mathematical symbols, including parentheses (·), square brackets [·], and curly braces {·}, signify vectors, sequences, and sets respectively. This distinction is maintained to ensure clarity of notation throughout the text.

**Definition 1: Information cascade**

A social network can be represented as a directed graph $G = (V, E)$, where $V$ symbolizes a set of nodes (users), and $E$ is a set of edges (user relationships). When a node $u$ is activated for the first time, it can potentially activate its currently inactive neighbor $v$ with a probability $p_{uv}$, depending on the edge $(u, v)$. An information cascade can thus be defined as a sequence of activations $(u_1, u_2, \cdots, u_n)$ where each user $u_i$ (for $i$ = 2, 3, ..., $n$) is activated by a previous user $u_j$ with $j < i$ [37].

**Definition 2: Information cascade snapshot**

For a given cascade $p_i$, the propagation behavior observed at time $t_j$ within time span $T$ is denoted as $D_i(t_j) = \{(u_k^i, v_k^i, t_j) | u_k^i, v_k^i \in V, t_j \in [0, T), 1 \le k \le l_j\}$, where the tuple $(u_k^i, v_k^i, t_j)$ symbolizes the transmission of message $m_i$ from user $u_k$ to user $v_k$ at time $t_j$. Here, $l_j$ represents the aggregate of the propagation behaviors observed up to time $t_j$. Then, the cascade snapshot for $p_i$ at time $t_j$ can be denoted as $S(p_i) = [D_i(t_1), \cdots, D_i(t_j)]$ [33].

**Definition 3: Global graph**

A global graph refers to a comprehensive representation of a network that encapsulates all the nodes and their corresponding edges within the entire system [4, 15]. Unlike a subgraph or local graph that may focus on specific regions or clusters within the network, a global graph provides a complete view of the entire structure [38]. It is essential for a comprehensive understanding of structural dynamics and relational complexity.

**Definition 4: Information cascade prediction**

Given an observed information cascade represented as graph $C_i(t)$ within a given observation window $W$, the goal of information cascade prediction in this study is to estimate the increase in cascade size $\Delta R_w^i$ over a future time interval $\Delta t$. Formally, $\Delta R_w^i = R_{w+\Delta t}^i - R_w^i$. Here, $R_w^i$ represents the number of nodes (users) activated during the observation window, while $R_{w+\Delta t}^i$ denotes the total number of nodes activated after time interval $\Delta t$ [15].

# 4 METHODOLOGY

The CasCIFF model, presented in this research, fuses multi-source information based on a multi-task learning framework, which improves the cascade prediction performance. The model is specifically designed to perform two symbiotically connected tasks: user classification according to user influence and information cascade prediction using multi-source information. A schematic overview of the model architecture is illustrated in Fig. 1. The configuration of the model can be systematically divided into four core modules:

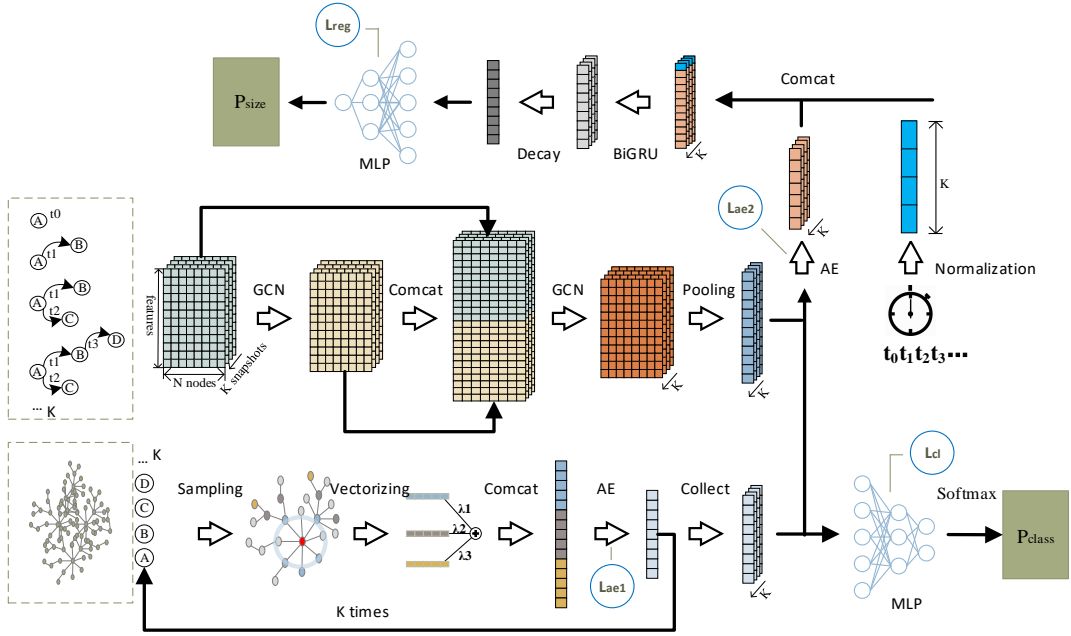

Fig. 1. Architecture of the CasCIFF model.

(1) User representation learning: In this module, a user's influence is measured by analyzing historical interactions between the user and their multi-hop neighbors in the underlying social network. Utilizing autoencoder techniques, principal component analysis is then executed to extract user features and improve computational efficiency.

(2) Cascade representation learning: This module captures essential features for information cascade prediction based on evolutionary trajectory and instantaneous velocity. The timing of user actions is taken into account to compute the weights of the adjacency matrix at each time point, allowing for a nuanced understanding of propagation patterns.

(3) User classifier: Here, a multi-layer perceptron (MLP) is used to distinguish opinion leaders from generic users according to user influence. This task provides an inherent constraint that allows the model to refine user characteristics with increased reliability, thereby improving the accuracy of cascade predictions.

(4) Information cascade predictor: The focus of this module is to integrate information about user influence, event timing, and cascade graphs. Using a Bi-direction Gated Recurrent Unit (BiGRU) [29], a subtype of recurrent neural networks, the module extracts spatiotemporal attributes from the amalgamated data. A time decay factor is then applied to weight a comprehensive vector for subsequent cascade prediction efforts.

## 4.1 User influence representation

The dynamics of information diffusion within social networks are primarily shaped by users' retweeting patterns. Nonetheless, these users exhibit considerable heterogeneity in their behaviors and influence. A particular subset of users, termed 'opinion leaders', exert considerable influence over the views, beliefs, and actions of others [39]. Traditional identification of influential users typically relies on centrality measures like degree centrality [40]. However, the advent of fake followers has undermined the reliability of these conventional methods [41].

In response to this predicament, we propose a global graph-based method that simultaneously weighs the quantity and quality of user connections. This method aggregates information from both direct neighbors and multi-hop neighbors throughout the network. Fig. 2 offers a visual illustration of this methodology, exemplified through the use of 3-hop neighbors.

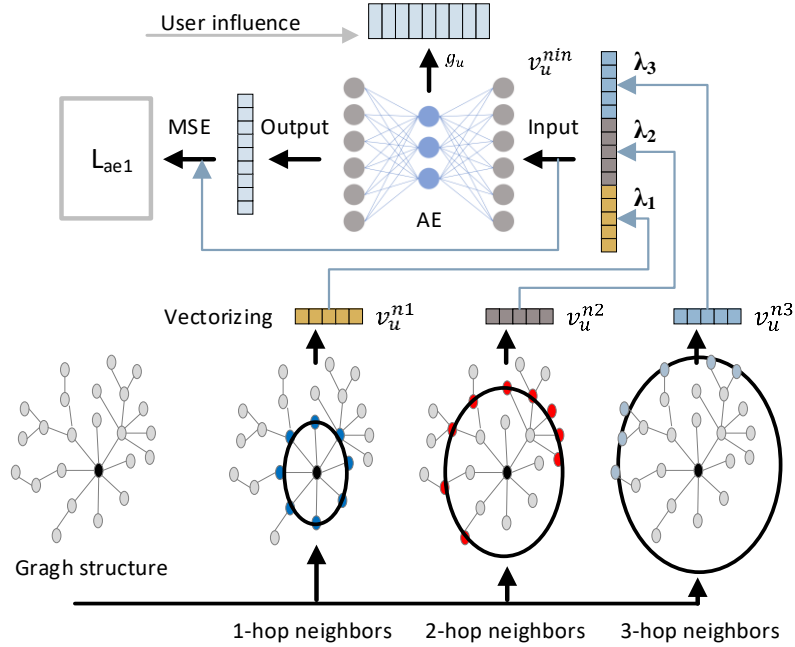

Fig. 2. Schematic illustration of representing user influence based on the global graph, exemplified using 3-hop neighbors. Note that for the sake of clarity, the vectorization is depicted only for a single user; nonetheless, the same procedure should be applied to all users in practice.

It should be noted that in real-world applications, the computational complexity of this approach grows exponentially with an increase in the order of hops, which may limit its usefulness. To mitigate this problem, the Hierarchical Random Sampling (HRS) technique [42] is introduced into the sampling process. This strategy is based on the idea that direct connections offer a more accurate reflection of a user's influence than distant connections. The procedure is described in detail as follows:

For a given user, $k$ nodes are randomly selected from the 1-hop neighbors to form a centered subgraph. Similarly, $k \times 2^{-1 \times (2-1)}$ nodes are selected from the 2-hop neighbors and $k \times 2^{-1 \times (n-1)}$ nodes from the *n*-hop neighbors. Note that all samples are drawn without replacement.

Let $v_u^{n1}, v_u^{n2}$ and $v_u^{ni}$ represent the 1-, 2-, and *i*-hop neighbors of user *u*, respectively. For each hop order, $v_u^{ni}$ is formed according to the top-*s* influential nodes with a good influence degree, which mitigates the interference from fake followers. In this investigation, s is fixed at 50. Recognizing that the importance of neighboring nodes may vary with hop distance, neighborhood information at various hop distances is aggregated with the different weights representing importance. To describe the user, the input data for user *u*, denoted as $v_u^{nin}$, can be mathematically expressed as:

$$v_u^{nin} = concat(\lambda_1 * v_u^{n1}, \lambda_2 * v_u^{n2}, \cdots, \lambda_i * v_u^{ni}). \quad (1)$$

Here, $\lambda$ denotes a learnable weight, and concat(·) indicates the concatenation operation.

Unfortunately, the input $v_u^{nin}$ just reflects the user's external performance rather than his intrinsic attributes. To bridge this gap, we employ an autoencoder (AE) approach based on the work of Feng et al.[17]. This process can be mathematically represented as:

$$g_u = \sigma\left(fc3\left(\sigma\left(fc2\left(\sigma\left(fc1(v_u^{nin})\right)\right)\right)\right)\right), \quad (2)$$

$$L_{ae1} = mse\left(v_u^{nin}, \sigma\left(fc5\left(\sigma(fc4(g_u))\right)\right)\right). \quad (3)$$

In these expressions, $\sigma(\cdot)$ represents a non-linear activation function like the ReLU function, *fc*(·) denotes a fully connected layer, and *mse*(·) symbolizes the mean squared error. Here, the variable $g_u$ corresponds to the global influence of user *u*, which is the optimal solution in the sense of minimizing the loss function $L_{ae1}$. The influence $g_u$ provides the basis for subsequent tasks such as user classification and cascade prediction.

## 4.2 Local cascade structure representation

The initial trajectory of information diffusion is critical in determining its eventual diffusion scale. Capturing information from early message diffusion is essential, but difficult for cascade prediction due to limited observations. Fortunately, both cascade structure and diffusion rate provide important clues for cascade prediction. To exploit these clues, numerous methods have been developed using the 'snapshot' technique.

To generate these snapshots, we adopt a sampling mechanism analogous to that described in CasCN [14]. As depicted in Fig. 3, the cascade graph $C_i(t)$ is sampled within a given time window *W* to yield a sequence of subgraphs, denoted as $G_i^T$, which is expressed as follows:

$$G_i^T = \{[g_i^{t_1}, g_i^{t_2}, \cdots, g_i^{t_{m-1}}] | t_j \in [0, T), j \in [1, m]\} \quad (4)$$

where $g_i^{t_j}$ is a subgraph, or a snapshot of the cascade, sampled from $C_i(t)$ at time $t_j$. Each subgraph $g_i^{t_j}$ is represented by an adjacency matrix $\boldsymbol{\alpha}_i^{t_j}$, where rows correspond to the node labels in alphabetical order from top to bottom and columns correspond to edges.

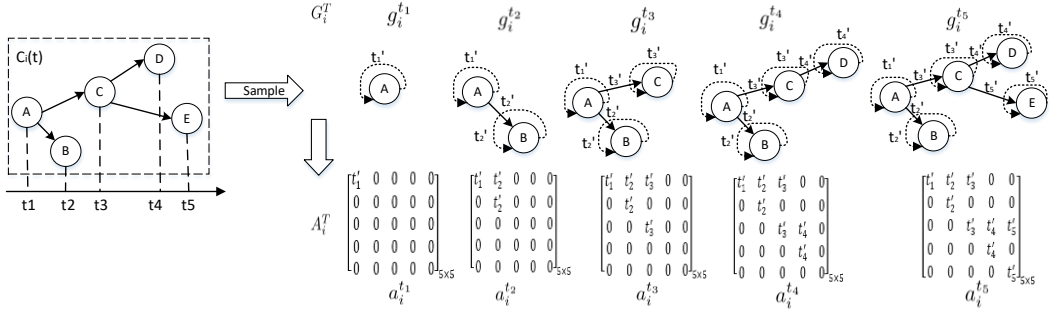

Fig. 3. The sampled subgraphs and the corresponding weighted adjacency matrices.

Unfortunately, the adjacency matrix employed by CasCN [14] only captures the retweeting relationships between nodes, neglecting the importance of the timing of actions. Furthermore, the irregular time intervals between neighboring snapshots prohibit the extraction of diffusion speed, which is crucial information. To overcome these limitations, we replace the original matrix with a weighted matrix. The weights in this context represent the standardized participation time, calculated by

$$t_i' = 1 - \frac{t_i}{T_o} \tag{5}$$

where $t_i$ denotes the participation time of the $i$-th user and $T_o$ is the observation time of the cascade. Obviously, (5) embodies the transmission rule that earlier retweets have a stronger impact on diffusion outcomes. In addition, we use the elements along the diagonal of the adjacency matrix to document self-links, thereby enhancing the rank of the matrix and potentially increasing the corresponding information entropy.

To extract more robust features from cascade snapshots, we employ a two-layer Graph Convolutional Network (GCN). The underlying computations are expressed as:

$$\boldsymbol{h} = \sigma\left(\mathbf{W}_1 *\mathcal{G}\left(\boldsymbol{\alpha}_i^{t_j}\right) + \mathbf{b}_1\right), \tag{6}$$

$$g_c = \text{pooling}\left(\sigma\left(\mathbf{W}_2 *\mathcal{G}\left(\text{concat}\left(\boldsymbol{h}, \boldsymbol{\alpha}_i^{t_j}\right)\right) + \mathbf{b}_2\right)\right). \tag{7}$$

Here, $*\mathcal{G}$ denotes the graph convolution operation, $\boldsymbol{\alpha}_i^{t_j}$ indicates the weighted adjacency matrix of the cascade snapshot $g_i^{t_j}$ at time $t_j$. Other parameters $\mathbf{W}_1$, $\mathbf{W}_2$, $\mathbf{b}_1$, and $\mathbf{b}_2$ are learnable weight matrices and bias terms. After a mean pooling operation, $g_c$ is then used as part of the input data for the prediction module.

## 4.3 Multimodal information fusion

In the process of information diffusion, user influence and cascade structure are the determinants of the diffusion scale. The former offers insight into the number of potential propagators, while the latter elucidates the pattern of information transmission. Recognizing the complex interdependence of these factors, we introduce a multimodal fusion scheme to exploit both micro- and macro-level information.

For a given information cascade, $C_i(t)$, the global influence of each participant is represented as $G_m = \{g_u | u \in V\}$ and the local cascade structure as $g_c$. To synthesize these representations, we employ an autoencoder for data dimensionality reduction. The resulting output and the loss function are

$$Stru_u = \sigma\left(\text{fc}\left(\sigma\left(\text{fc}\left(\sigma\left(\text{fc}(concat(g_u, g_c))\right)\right)\right)\right)\right), \quad (8)$$

$$L_{ae2} = mse\left(v_u^{neo}, \sigma\left(\text{fc}\left(\sigma(\text{fc}(v_u^{neo}))\right)\right)\right). \quad (9)$$

Then, a timestamp is appended to the output vector, $Stru_u$, to capture temporal nuances, which is especially important for prediction accuracy when the vector is fed into the recurrent neural network. This process can be mathematically expressed as

$$Stru\_t_i = \{concat(Stru_u, t_u')| u \in V\}. \quad (10)$$

It should be noted that the time used here is still the standardized participation time, calculated as described in (5).

## 4.4 Multi-task learning

### 4.4.1 Information cascade prediction

The analysis of information cascades in both chronological and reverse chronological order offers valuable insights into the inherent relationships between nodes. In line with previous research [15, 29], we utilize a BiGRU in the prediction module. At each step *i*, with the fusion information $Stru\_t_i$ and the previous hidden state $h_{i-1}$ as inputs, the BiGRU updates the state $h_i$ in both forward and backward directions. Then, the final output of BiGRU is a concatenation of both forward and backward hidden vectors, which would be used as input data for the prediction task module. This process can be mathematically expressed as

$$\vec{h}_i = GRU_{fwd}(Stru_{t_i}, \vec{h}_{i-1}), \quad (11)$$

$$\overleftarrow{h}_i = GRU_{fwd}(Stru_{t_i}, \overleftarrow{h}_{i-1}), \quad (12)$$

$$h_i = concat(\vec{h}_i, \overleftarrow{h}_i). \quad (13)$$

Time decays, introduced by various factors, can greatly influence the cascade dynamics, especially in predictive contexts. To improve predictive accuracy, we incorporate a time-decay

mechanism based on a non-parametric methodology as advocated by Cao et al [13]:

$$h'_i = \lambda_m h_i. \tag{14}$$

Here, $\lambda_m$ is a learnable discrete variable. If the time window $W$ is divided into $L$ disjoint intervals, $\lambda_m$ denotes the weight parameter corresponding to the $m$-th interval that $h_i$ falls into.

Then, the increase of the cascade size after a fixed time interval $T$ is predicted through a Multi-Layer Perceptron (MLP):

$$\Delta R_T^i = \text{MLP}_{reg}(h'_i). \tag{15}$$

The loss function of the MLP can be defined as follows:

$$L_{reg} = \frac{1}{M} \sum_{i=1}^{M} (\log_2 \Delta R_T^i - \log_2 \Delta \hat{R}_T^i)^2, \tag{16}$$

where $\Delta R_T^i$ is the actual increase in cascade size, $\Delta \hat{R}_T^i$ is the predicted output, and $M$ is the total number of cascades.

### 4.4.2 User classification

Because of the difference between ordinary users and opinion leaders in the power of information propagation, user classification is helpful for cascade prediction. For this purpose, user influence must be represented in advance by

$$P(u|g_u) = softmax(MLP_{cl}(g_u)). \tag{17}$$

The loss function for the user classification task, measured by cross-entropy, is:

$$L_{cl}(\hat{P}, P) = \frac{1}{N} \sum_{i=1}^{N} cross\_entropy(\hat{P}, P), \tag{18}$$

where $N$ is the total number of users in the training dataset.

Finally, the combined loss function for the CasCIFF model is

$$L = L_{reg} + L_{cl} + L_{ae1} + L_{ae2} + L_{rgl} \tag{19}$$

Here, $L_{rgl}$ denotes the L2-norm regularization term over all parameters to mitigate overfitting, calculated by the sum of the squares of all weights of the model, i.e.,

$$L_{rgl} = \sum \|w\|^2. \tag{20}$$

The calculation formulas for the remaining terms on the right-hand side of (19) have already been given before.

# 5 EXPERIMENT SETUPS

## 5.1 Datasets

To evaluate the CasCIFF model, we used three publicly available datasets: Weibo, Twitter, and APS. These datasets, which have been widely used in previous research [10, 13, 15, 17, 34], include information about cascade ID, posting time, posting user, and retweet paths. These datasets are introduced in detail below.

Sina Weibo: This dataset, provided by Cao et al. [13], can be accessed publicly at https://github.com/CaoQi92/DeepHawkes. It encompasses Weibo messages that were posted and retweeted more than 10 times on June 1, 2016. Because most cascades reached saturation within the first 24 hours after posting [13], the cascade size within this time frame is treated as the final prediction target. The observation time $T$ was set at 0.5 and 1 hour, respectively.

Twitter: This dataset, collected by Weng et al.[43], is publicly available at https://carl.cs.indiana.edu/data/#virality2013. It contains public tweets written in English and posted between March 24 and April 25, 2012. The final prediction target was the number of retweets within 32 days, and the observation time $T$ was set at 1 and 2 days.

APS: This dataset, re-organized by the American Physical Society (APS), is available at https://journals.aps.org/datasets and comprises all papers published in 17 APS journals from 1893 to 2017 [12]. We only chose the papers published before 1997, so that each paper had a 20-year history. The total number of citations in 20 years served as the final prediction target. The observation time $T$ was set at 3 and 5 years.

Similar to the experiment conducted by Xu [15], all cascades must comprise more than 10 nodes. In order to enhance the level of challenge, for graphs with more than 100 nodes, only the first 100 nodes (ordered by adoption time) were selected. For all datasets, 70% of the randomly shuffled cascades were used for training, 15% for validation, and the remaining 15% for testing. The data used for the experiments are shown statistically in Table 1. Note that the final data used is less than the original, as unqualified data were discarded.

Table 1. Descriptive statistics of three datasets.

|  | Data sets | Sina Weibo | | Twitter | | APS | |
|---|---|---|---|---|---|---|---|
| Posts | All | 119313 | | 88440 | | 207685 | |
| Nodes | All | 6738040 | | 490474 | | 616316 | |
| Edges | All | 15249636 | | 1903230 | | 3304400 | |
| Time | - | 0.5h | 1h | 1day | 2day | 3year | 5year |
| Posts | train | 21425 | 29863 | 9639 | 12739 | 18511 | 32102 |
|  | valid | 4590 | 6398 | 2066 | 2730 | 3967 | 6879 |
|  | test | 4592 | 6395 | 2065 | 2729 | 3966 | 6879 |

## 5.2 Baselines

In order to assess the effectiveness of the proposed method, we compared it with several state-of-the-art models:

Feature-based: In this method, an MLP is trained using features extracted from information cascades. These features fall into three categories: user attributes (represented by the number of followers), structural properties (including average degree, number of leaf nodes, and depth), and temporal properties (such as average, maximum, and minimum response time).

DeepHawkes [13]: An innovative model that emulates Hawkes processes using a deep learning framework. It incorporates a non-parametric time-decay method to account for temporal dynamics.

CasCN [14]: The first model uses graph representation learning to overcome the limitations of spectral graph convolution for directed graphs.

TCN [44]: This model relies only on temporal features for prediction, eliminating the need for structural features of cascades.

VaCas [45]: The first approach based on Bayesian learning to cascade prediction uses pre-trained cascade node embeddings as input.

MuCas [10]: This model uses a multi-scale graph capsule network and an influence attention mechanism to learn the latent representation of cascade graphs.

CasFlow [15]: An advanced model built upon VaCas enhances the predictive power of the model with global user representation and regularization flow modules.

## 5.3 Evaluation metrics

Consistent with previous research [12, 15, 34], our evaluation employs two widely recognized metrics: Mean Square Logarithmic Error (MSLE) and Mean Absolute Percentage Error (MAPE). They are mathematically defined as:

$$\text{MSLE} = \frac{1}{n}\sum_{i=1}^{n}(\log_2(y_i + 1) - \log_2(\hat{y}_i + 1))^2 \tag{21}$$

$$\text{MAPE} = \frac{1}{n}\sum_{i=1}^{n}\left|\frac{\log_2(y_i + 1) - \log_2(\hat{y}_i + 1)}{\log_2(y_i + 1)}\right| \tag{22}$$

In these equations, $n$ stands for the total number of cascades. $\hat{y}_i$ denotes the predicted incremental popularity from our model, while $y_i$ corresponds to the actual observed value (ground truth).

## 5.4 Parameter configuration

To compare all methods fairly, we set parameters identical to those in DeepHawkes [13] and CasFlow[15] to obtain optimal performance. Specifically, the time interval for the non-parametric time decay effect is set to three hours for Twitter and three months for APS. For Sina Weibo, we adjust the time interval to five minutes and ten minutes when the length of the time window is 0.5 hours and 1 hour, respectively.

The learning rate is iteratively explored within $10^{\{-1,\cdots,-5\}}$. The regularization coefficients (L1 or L2) are sampled from $10^{\{0,\cdots,-8\}}$. The batch size is fixed at 64, and all remaining hyper-parameters follow those specified in the original papers of each baseline.

All models are trained using the Adam optimizer [46], with early stopping implemented when validation errors do not decrease for ten consecutive epochs.

# 6 EXPERIMENTAL RESULTS

## 6.1 Performance analysis

The experiments were conducted on a computer platform equipped with an NVIDIA GeForce RTX3080Ti, an Intel i9-10920X processor, and 64 GB of memory. The performance comparison between CasCIFF and the baselines is systematically presented in Table 2.

As can be seen in Table 2, CasCIFF consistently outperforms its closest competitor, CasFlow, in terms of performance metrics. The only deviation from this trend is observed when the time window is set to 0.5 hours for the Weibo dataset, considering the MAPE metric. Such consistent outperformance underscores the effectiveness and robustness of the proposed model framework.

Table 2. Performance comparison in term of MSLE and MAPE for all methods.

| Model | Twitter | | | | Weibo | | | | APS | | | |
|---|---|---|---|---|---|---|---|---|---|---|---|---|
| | 1day | | 2day | | 0.5h | | 1h | | 3year | | 5year | |
| | MSLE | MAPE | MSLE | MAPE | MSLE | MAPE | MSLE | MAPE | MSLE | MAPE | MSLE | MAPE |
| Feature-Deep | 7.551 | 0.509 | 6.322 | 0.428 | 2.828 | 0.248 | 2.743 | 0.277 | 1.849 | 0.272 | 1.736 | 0.295 |
| DeepHawkes | 7.216 | 0.587 | 5.788 | 0.536 | 2.891 | 0.268 | 2.796 | 0.282 | 1.573 | 0.271 | 1.324 | 0.335 |
| CasCN | 7.183 | 0.547 | 5.561 | 0.525 | 2.804 | 0.254 | 2.732 | 0.273 | 1.562 | 0.286 | 1.421 | 0.265 |
| TCN | 7.104 | 0.477 | 5.072 | 0.383 | 2.608 | 0.234 | 2.517 | 0.265 | 1.810 | 0.267 | 1.684 | 0.285 |
| VaCas | 7.228 | 0.494 | 5.354 | 0.373 | 2.524 | 0.227 | 2.507 | 0.255 | 1.806 | 0.267 | 1.673 | 0.284 |
| MuCas | 6.944 | 0.506 | 5.334 | 0.442 | 2.819 | 0.260 | 2.650 | 0.272 | 1.552 | 0.236 | 1.399 | 0.247 |
| CasFlow | 6.954 | 0.455 | 5.143 | 0.361 | 2.402 | **0.210** | 2.279 | 0.238 | 1.361 | 0.222 | 1.354 | 0.248 |
| CasCIFF (improves) | **6.805** ↑2.2% | **0.452** ↑0.6% | **5.024** ↑2.4% | **0.358** ↑0.8% | **2.334** ↑2.9% | 0.220 ↓4.5% | **2.263** ↑0.7% | **0.236** ↑0.8% | **1.133** ↑20.3% | **0.206** ↑7.8% | **1.233** ↑9.8% | **0.238** ↑4.2% |

Note that a lower index value indicates better performance. For clarity and ease of reference, the optimal values across scenarios are highlighted and colored red in the table. In its current configuration, CasCIFF limits its sampling to nodes within the 2-hop neighborhood. However, by tailoring the sampling approach to specific datasets, there is potential to further optimize CasCIFF's performance.

It is worth noting that information diffusion as a human activity is much more complex than mechanical movement. Although from a statistical point of view, long-term and large sample observations should obey the essential laws of information propagation, the individual cascade evolution process, especially for the ultra-short-term results, is subject to the influence of participant individuality and often seriously deviates from the normal situation. If the participants have not been seen in the training set, their personalities are not known, which makes cascade prediction difficult or results unreliable. Only 70% of the samples in the benchmark are used for training, so there is inevitably a large number of "strangers" in the data used for evaluation, making cascade prediction extremely challenging. Although this benchmark can test the superiority of the algorithms, it is unlikely that any improvement in the algorithm will result in a significant increase in performance metrics. Perhaps increasing the percentage of training samples could alleviate this problem, but the use of a common benchmark facilitates the reproduction of the results of the baseline algorithms and ensures the correctness of the implemented code. For this reason, the same benchmark as in other literature [12, 15, 17, 34] is used in this work.

## 6.2 Complexity analysis

In order to gain a deeper understanding of the advantages and limitations, we compared CasCIFF with the baselines in terms of number of parameters and training time. The detailed comparison is presented in Table 3. With respect to number of parameters, DeepHawkes has the highest requirement, around 250M. This is due to the need to embed data for all users within the social network. The total number of parameters for this task can be calculated as $N \times F_{\text{dim}}$ parameters, where $N$ is the total number of users, and $F_{\text{dim}}$ signifies the embedding dimension.

Table 3. Model parameters and computation time (in seconds) for Twitter, Weibo and APS.

| Methods | Para. Size | Time cost per epoch (s) | | | | | |
|---|---|---|---|---|---|---|---|
| | | Tw. (1d) | Tw. (2d) | Wb. (0.5h) | Wb. (1h) | APS (3y) | APS (5y) |
| Feature | 63B | 1 | 1 | 1 | 1 | 1 | 1 |
| DeepH. | 250M | 69 | 90 | 198 | 301 | 104 | 202 |
| CasCN | 278K | 794 | 1121 | 1800 | 2867 | 1420 | 3048 |
| TCN | 9K | 2 | 3 | 4 | 7 | 3 | 7 |
| VaCas | 310K | 15 | 18 | 32 | 46 | 26 | 49 |
| MuCas | 495K | 110 | 146 | 250 | 305 | 201 | 352 |
| CasFlow | 1M | 22 | 28 | 47 | 68 | 42 | 73 |
| C.CIFF_n1 | 601K | 82 | 114 | 181 | 237 | 167 | 290 |
| C.CIFF_n2 | 601K | 84 | 116 | 183 | 254 | 176 | 293 |
| C.CIFF_n3 | 601K | 88 | 118 | 186 | 263 | 181 | 297 |
| C.CIFF_n4 | 601K | 94 | 126 | 197 | 270 | 183 | 301 |
| C.CIFF_n5 | 601K | 102 | 131 | 204 | 285 | 185 | 307 |

Note that some figures after the decimal point have been omitted due to space limitations. Here Tw. and Wb. refer to Twitter and Weibo respectively.

When it comes to training time, CasCN emerges as the most resource-intensive, consuming about 50 minutes per epoch for APS when T=5y. CasCIFF shows a balanced profile with a moderate number of parameters and training time. This suggests a significant potential for computational improvement, which we intend to explore in subsequent research efforts.

## 6.3 Ablation study

To assess the contribution of each component within the CasCIFF framework, we conducted an ablation study. For this purpose, we present five derived versions of the CasCIFF model:

CasCIFF-Local, CasCIFF-Global, CasCIFF-Time, CasCIFF-Decay, and CasCIFF-Class. The specific modifications to each version are described below:

(1) CasCIFF-Local omits the local representation module.

(2) CasCIFF-Global excludes the global cascade representation module.

(3) CasCIFF-Time removes all time-related features, including the time decay module and the weighted cascade snapshot.

(4) CasCIFF-Decay specifically removes the time decay module.

(5) CasCIFF-Class discards the user identity classifier.

Table 4 delineates a performance comparison between the primary CasCIFF model and its derived versions. It is evident that the original CasCIFF model generally achieves superior results compared to its variants, highlighting the importance and effectiveness of each integrated component within the framework.

Table 4. Performance comparison between CasCIFF and five variations.

| Model | Twitter | | | | Weibo | | | | APS | | | |
|---|---|---|---|---|---|---|---|---|---|---|---|---|
| | 1day | | 2day | | 0.5h | | 1h | | 3year | | 5year | |
| | MSLE | MAPE | MSLE | MAPE | MSLE | MAPE | MSLE | MAPE | MSLE | MAPE | MSLE | MAPE |
| CasCIFF-Local | 6.838 | 0.466 | 5.154 | 0.368 | 2.651 | 0.248 | 2.572 | 0.259 | 1.809 | 0.267 | 1.676 | 0.285 |
| CasCIFF-Global | 7.290 | 0.494 | 5.554 | 0.414 | 2.424 | 0.230 | 2.361 | 0.245 | 1.149 | 0.207 | 1.240 | 0.240 |
| CasCIFF-Time | 11.47 | 0.807 | 10.29 | 0.757 | 2.825 | 0.258 | 3.889 | 0.342 | 1.300 | 0.226 | 1.505 | 0.271 |
| CasCIFF-Decay | 6.928 | 0.481 | 5.005 | 0.365 | **2.278** | 0.226 | 2.224 | **0.233** | 1.126 | 0.205 | 1.256 | 0.238 |
| CasCIFF-Class | **6.649** | 0.463 | 4.910 | 0.378 | 2.318 | 0.227 | 2.225 | 0.241 | 1.110 | 0.203 | 1.265 | 0.240 |
| CasCIFF_n1 | 7.002 | 0.453 | 5.114 | 0.399 | 2.375 | 0.227 | 2.314 | 0.242 | 1.189 | 0.212 | 1.283 | 0.245 |
| CasCIFF_n2 | 6.805 | **0.452** | 5.024 | 0.358 | 2.334 | 0.220 | 2.263 | 0.236 | 1.133 | 0.206 | **1.233** | **0.238** |
| CasCIFF_n3 | 6.763 | 0.503 | 4.962 | 0.39 | 2.327 | **0.217** | **2.219** | 0.242 | 1.134 | 0.209 | 1.259 | **0.238** |
| CasCIFF_n4 | 6.787 | 0.51 | **4.869** | **0.356** | 2.432 | 0.234 | 2.242 | 0.241 | **1.104** | **0.200** | 1.243 | 0.239 |
| CasCIFF_n5 | 6.795 | 0.501 | 4.911 | 0.363 | 2.451 | 0.235 | 2.265 | 0.236 | 1.11 | 0.207 | 1.268 | 0.242 |

Note that a lower index value indicates better performance. For clarity, the optimum performance in each case is highlighted in bold red typeface in the table.

Interestingly, the CasCIFF-Decay version shows commendable performance, especially on the Weibo dataset. This phenomenon could be attributed to the relatively short time window in which the effects of time decay are less pronounced.

Furthermore, the CasCIFF-Class version exhibits significant performance on the Twitter dataset with a 1-day time window. Such an observation may indicate that globally influential users, such as opinion leaders, don't always have a significant impact on a particular information spread, leading to variations in performance across different contexts.

It is also observed that the influence representation of users from neighbors at different distances can influence the predictive accuracy of information cascades. A more detailed discussion about user influence is provided in the following section.

# 7 DISCUSSION AND CONCLUSION

## 7.1 Representation of user influence

In social networks, weak ties (also termed distant connections) and strong ties (or close connections) play distinct roles in information diffusion. Specifically, strong ties predominantly influence individual behaviors and attitudes, while weak ties primarily assist in the acquisition of new information [47]. Consequently, users' influence in these networks is not limited to immediate connections (1-hop neighbors) but can extend to connections of these connections (2-hop neighbors) and potentially extend even further.

From this perspective, one might reasonably hypothesize that a broader scope of neighborhood information would provide a more accurate representation of user influence. Nonetheless, according to small-world network theory, which suggests that the average path length between individuals is surprisingly short, approximately 5.2 intermediaries [48], one could argue that the optimal representation of user influence should include the information of neighbors ranging from 1 to 5 hops. Incorporating information from more distant nodes, however, could introduce information redundancy and computational complexity.

To verify this hypothesis, experiments were conducted on three different datasets, with their respective results illustrated in Fig. 4. These results suggest that the role of neighbors with different hop distances varies across network platforms and observation windows. Contrary to our expectations, optimal MSLE values were typically achieved at the distance ranging from 2 to 4 hops, rather than at the 5-hop distance.

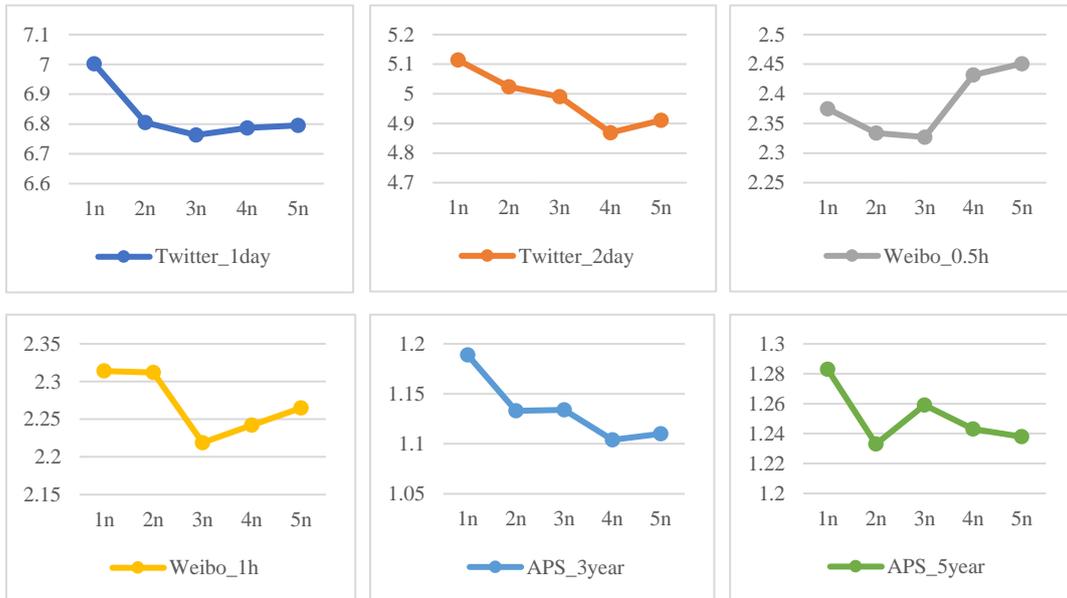

Fig. 4. The impact of neighborhood information with various hop distance on prediction accuracy of information cascade. Due to the correlation between MSLE and MAPE, only MSLEs are shown here.

This deviation from the hypothesis could stem from the noise introduced by distant network neighbors. These more distant connections, which typically indicate weaker ties, may not provide reliable insights into the target individual's behaviors or attitudes. This, in turn, could compromise their predictive utility in the task of cascade prediction.

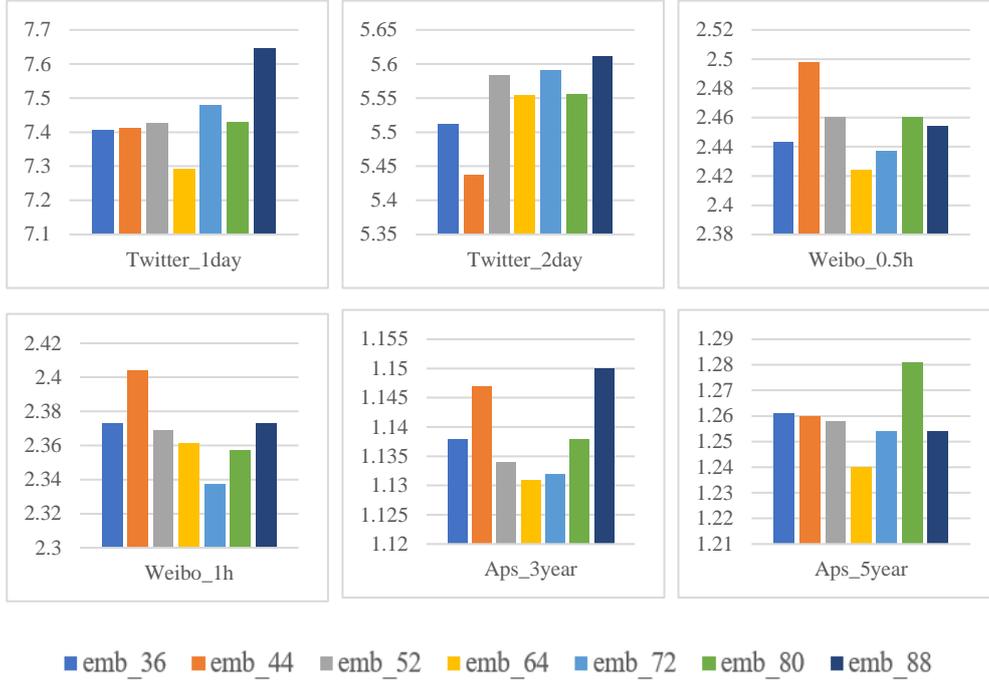

Fig. 5. The impact of imbedding dimension of users' influence on prediction accuracy of information cascade. Due to the correlation between MSLE and MAPE, only MSLE is shown here.

Additionally, we investigated the effect of data dimensionality on the accuracy of cascade prediction. In general, higher data dimensionality correlates with increased expressiveness. To determine whether higher dimensionality means better performance, we experimentally analyzed the role of vector dimensionality in determining prediction accuracy. The results, portrayed in Fig. 5, illustrate that the effect of dimensionality varies with the social network platform and the time window size. It is readily seen from Fig. 5 that vectors with 64 dimensions, as opposed to 88 dimensions, exhibited superior performance. Consequently, this research represents user influence using 64-dimensional vectors. Unexpectedly, these findings don't necessarily support the above hypothesis. The dimensionality of a vector representation may be related to the dimensionality of the features of the represented object. Excessive dimensionality introduces noise, and insufficient dimensionality leads to loss of information.

## 7.2 Model representation visualization

In order to comprehensively evaluate the representational ability of our model, we adopted a visualization strategy consistent with methods presented in previous research [17, 21, 28, 36]. We aimed to visually depict the latent features generated by our model, CasCIFF. Specifically, each cascade embedding $g'_u$ generated by CasCIFF is transformed into a two-dimensional space using the t-Distributed Stochastic Neighbor Embedding (t-SNE) method [49]. This transformation ensures that cascade networks with similar vector representations in the original space remain closely clustered in the two-dimensional representation.

Subsequently, we utilized a color-coding scheme for each data point that indicates the two-dimensional representation of an information cascade. By mapping specific feature values to distinct colors, the resulting visualization provides insight into the relationships between cascade representations and feature characteristics. The results are shown in Fig. 6, where each point represents a cascade in the test set (cascades with similar latent vectors are close in the plot) and the color of the point indicates a kind of feature. The darker the point, the higher the value of popularity, the number of nodes (NNodes) in the time window, the number of leaf nodes (NLNodes), mean reaction time (MRTime), or the number of opinion leaders (NOLeaders). It is clear that nodes within the same cluster share relatively similar brightness, indicating proximity in their characteristics.

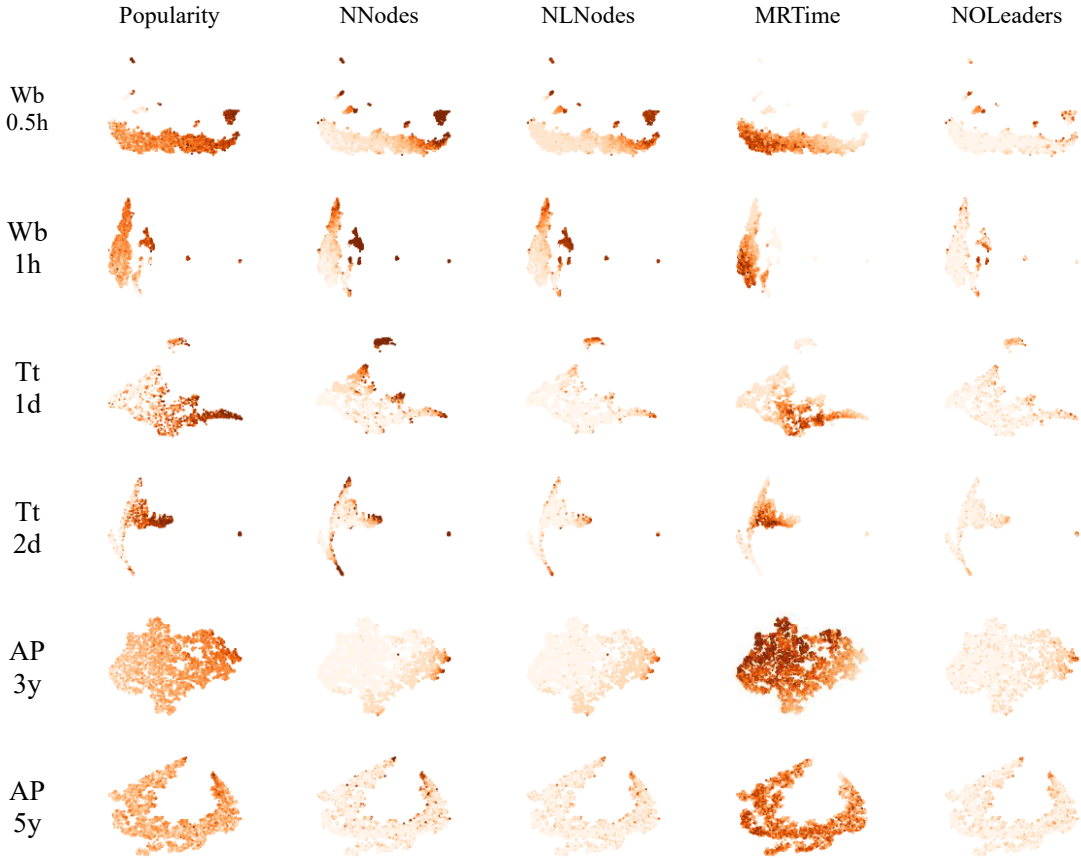

Fig. 6. Visualization of the representations of latent features using t-SNE in the different time windows on Weibo (Wb), Twitter (Tt), and APS (AP). The darker the point, the larger the value of popularity, the number of nodes (NNodes) in the time window, the number of leaf nodes (NLNodes), mean reaction time (MRTime), or the number of opinion leaders (NOLeaders).

## 7.3 Conclusion

In this study, we introduced a novel cascade prediction framework CasCIFF: Cross-Domain Information Fusion Framework specially designed for cascade prediction. This framework integrates users' global influence, local structural attributes, and temporal dynamics to simultaneously perform user classification and cascade prediction tasks.

To address the challenges posed by fake followers which can distort user influence, we developed a feature extraction algorithm based on multi-hop neighborhood information. Our investigations revealed that the optimal hop distance for user influence representation varies with network platforms and observation periods. Moreover, while globally influential users, such as opinion leaders, play a central role in many scenarios, their influence may not always be pronounced in specific information propagation events. Additionally, it also became evident that higher dimensionality of feature vectors does not guarantee better performance.

In order to capture information from early message propagation, we proposed a weighted adjacency matrix to describe the information cascade. Furthermore, a standardized participation time is appended to the integrated information so that the trend of information diffusion can be easily identified. We found that time decay effects become negligible in scenarios with relatively short time windows.

To integrate information from diverse sources, we constructed a deep learning-based framework. This framework employs an autoencoder for dimensionality reduction and MLPs for two symbiotically related tasks: user classification and information cascade prediction. Rigorous experimental evaluations on three publicly available datasets confirm that CasCIFF outperforms existing baselines in terms of MSLE and MAPE.

# ACKNOWLEDGMENTS

This work was supported by the National Social Science Foundation of China (20BXW097).